\documentclass[
  ,final            
  ]
  {aipproc}

\layoutstyle{6x9}

\newcommand{\pt}{\ensuremath{p_{\mathrm{T}}}}

\begin{document}

\title[Central Diffraction in ALICE]{Central Diffraction in Proton-Proton Collisions at $\sqrt{s}=7$\,TeV with ALICE at LHC}


\author{Felix Reidt\\ for the ALICE Collaboration}{
  address={Physikalisches Institut, Im Neuenheimer Feld 226, 69120 Heidelberg, Germany}
}

\begin{abstract}
A double-gap topology is used for filtering central-diffractive events
from a proton-proton minimum-bias data sample at a centre-of-mass
energy \mbox{$\sqrt{s}=7$\,TeV}. This topology is defined by
particle activity in the ALICE central barrel and absence of particle
activity outside. The fraction of events satisfying the double-gap
requirement $R_{DG}$ is found to be
\mbox{$7.63\pm0.02(stat.)\pm0.87(syst.)\times 10^{-4}$}. The
background of this double-gap fraction is estimated by studying the
contributions of non-diffractive, single- and double-diffractive
dissociation processes as modelled by Monte~Carlo event generators,
and is found to be about $10\%$.
\end{abstract}

\maketitle


\section{Introduction}
The ALICE experiment~\cite{Aamodt2008} at the Large Hadron Collider
(LHC) consists of the central barrel in the pseudorapidity range of
$-0.9<\eta<0.9$ and a muon spectrometer at
\mbox{$-2.5<\eta<-4.0$}. Additional detectors for triggering and event
classification are available in the range $-3.7<\eta<5.1$. Double-gap
events can be identified by requiring activity in the central barrel
and by the absence of activity in the more forward regions.
At LHC energies, about $25$\,mb, i.e.\ one third of the inelastic
hadronic cross section, are of diffractive
origin as e.g.\ measured by ALICE in a previous
study~\cite{Abelev2012}. These diffractive events are mediated by a
Pomeron~\cite{Donnachie2005} exchange. Central-diffractive events show
the characteristic double-gap topology due to their double-Pomeron
exchange. The central-diffractive cross section is predicted to be
about $800$\,$\mu$b~\cite{Ciesielski2012}.

\section{ALICE Detectors}
This analysis focuses on the central barrel of ALICE. The innermost
detector is the Inner Tracking System (ITS) which consists of six
layers. The first two layers are Silicon Pixel Detectors (SPD)
covering $-2.0<\eta<2.0$ and \mbox{$-1.4<\eta<1.4$}. The ITS is used
for vertex reconstruction and, together with the Time-Projection
Chamber (TPC), for tracking. In the forward region, the VZERO
scintillator hodoscope for trigger purposes is located at
$-3.7<\eta<-1.7$ and $2.8<\eta<5.1$. In addition multiplicity-density
measurements can be performed by the Forward-Multiplicity Detector
(FMD) situated at $-3.4<\eta<-1.7$ and $1.7<\eta<5.0$. It is a silicon
strip detector built for multiplicity density measurements. The low
material budget and low magnetic field of $0.5$\,T lead to a low
single particle \pt-threshold of 17\,MeV$/c$ for $50\%$ detection
efficiency. Further details on the ALICE detector system can be found
in ~\cite{Aamodt2008}.

\section{Double-Gap Topology}
\begin{figure}[h]
  \includegraphics[height=.27\textheight]{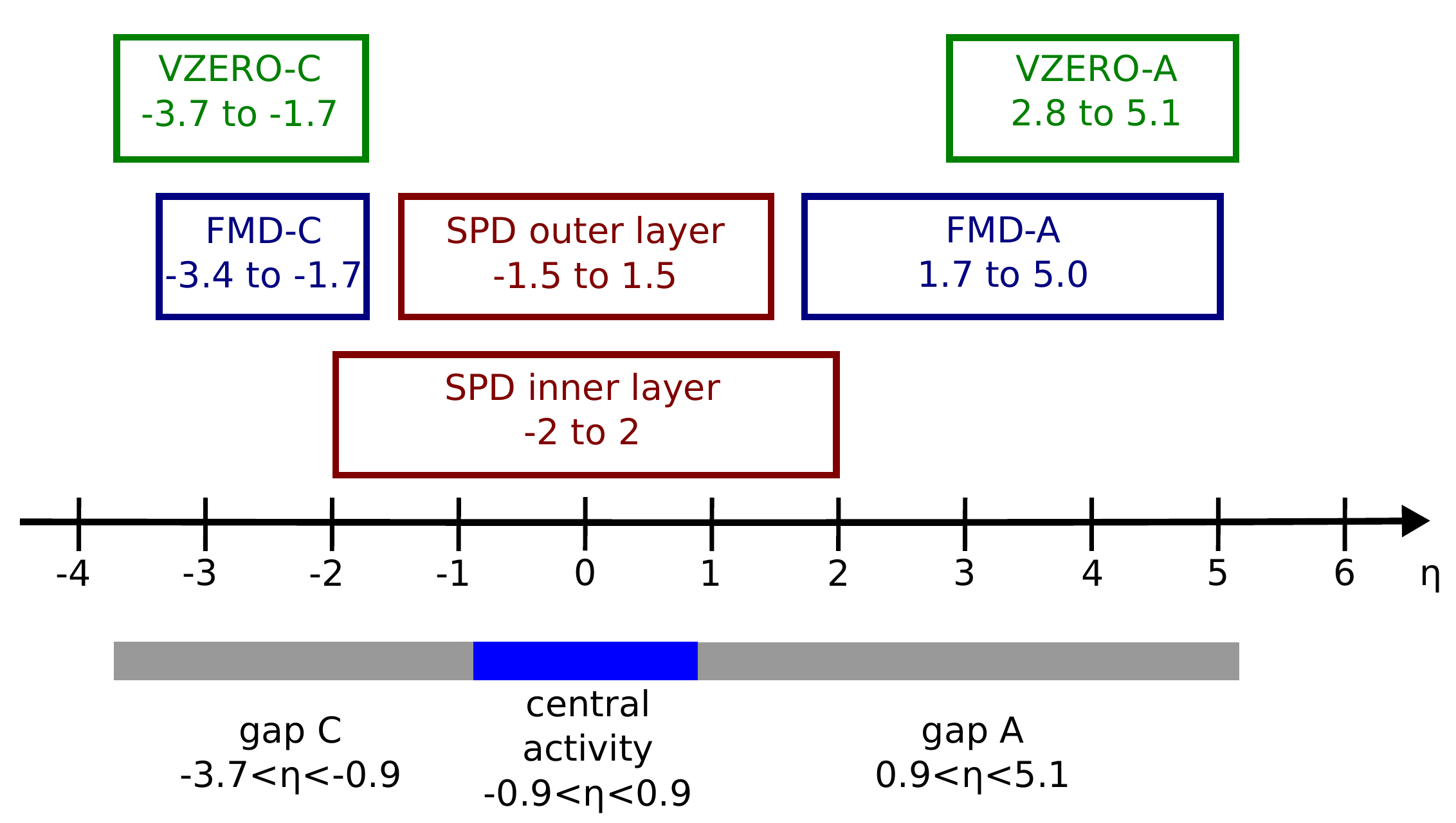}
  \caption{ALICE $\eta$ coverage by individual detectors used
    in this study and double-gap topology.}
  \label{fig:coverage}
\end{figure}

Central diffraction results from the two incoming protons interacting
via double-Pomeron fusion. The centrally produced system is
accompanied by large rapidity gaps. In central diffractive processes,
the protons can either stay intact or break up. The double-gap
topology can be exploited as filter for central diffraction. However,
at topology level, Pomeron, Reggeon or photon exchanges cannot be
distinguished. The contribution from Reggeon exchanges is assumed to
be comparable to lower energies, while the Pomeron contribution is
expected to increase with energy. The ALICE pseudorapidity coverage
and the corresponding gap definitions are shown in
Fig.~\ref{fig:coverage}. The gaps are defined by the absence of
activity induced by the passage of particles in the trigger signals of
VZERO, FMD and SPD in the pseudorapidity ranges of $-3.7<\eta<-0.9$
and $0.9<\eta<5.1$. The central activity is determined using SPD in
the range of $-0.9<\eta<0.9$. This double-gap configuration is a
trade-off of large gaps suppressing background and a reasonable
coverage for the centrally produced signal.

\section{Analysis}
The double-gap analysis is based on minimum-bias data at
$\sqrt{s}=7$\,TeV taken during the 2010 run. This data set was
acquired at low luminosity and low event pileup. The minimum-bias
trigger is the logical-OR of the two VZERO and the SPD trigger signal,
called MB$_\mathrm{or}$. A total of 300 million MB$_\mathrm{or}$
triggers are used for this analysis. Luminosity determination within
this data sample is possible by using the minimum-bias cross section
$\sigma_\mathrm{MBand}$ available from van-der-Meer scans and the
MB$_\mathrm{and}$ trigger count. The MB$_\mathrm{and}$ trigger is
derived from the coincidence of the two VZERO
subdetectors. Furthermore, a monitoring of the gap-fraction for all
involved detectors is done on a run-by-run basis. Runs deviating by
more than $3\sigma$ from the data-taking period mean are rejected. In
addition to the run selection, an event selection is
applied. Furthermore, a vertex within $\pm4$\,cm of the interaction
point in direction of the beam axis is required to ensure a continuous
gap coverage. Moreover, pileup is rejected by identifying multiple
vertices reconstructed from the SPD information. The number of
double-gap events $N_{DG}$ and the number of MB$_\mathrm{AND}$ events
$N_\mathrm{MBand}$ are analysed on a run-by-run basis.

First data-driven systematic-error estimates are available. The event
selection can influence the ratio of $N_{DG}$ to $N_\mathrm{AND}$, due
to the vertex requirement. For example, double-diffractive
dissociative events do not necessarily contain a vertex. This effect
is estimated to be about $6\%$. Furthermore, the central activity can
be missed, leading to a bias to lower cross section of approximately
$5\%$. On the contrary, missing a particle in the gap shifts the
cross section about $5\%$ towards higher values. The uncorrelated
error contribution is estimated to be $1.4\%$ from the spread of the
data-taking period means.

\section{Double-Gap Fraction}
\begin{figure}[h]
  \includegraphics[height=.3\textheight]{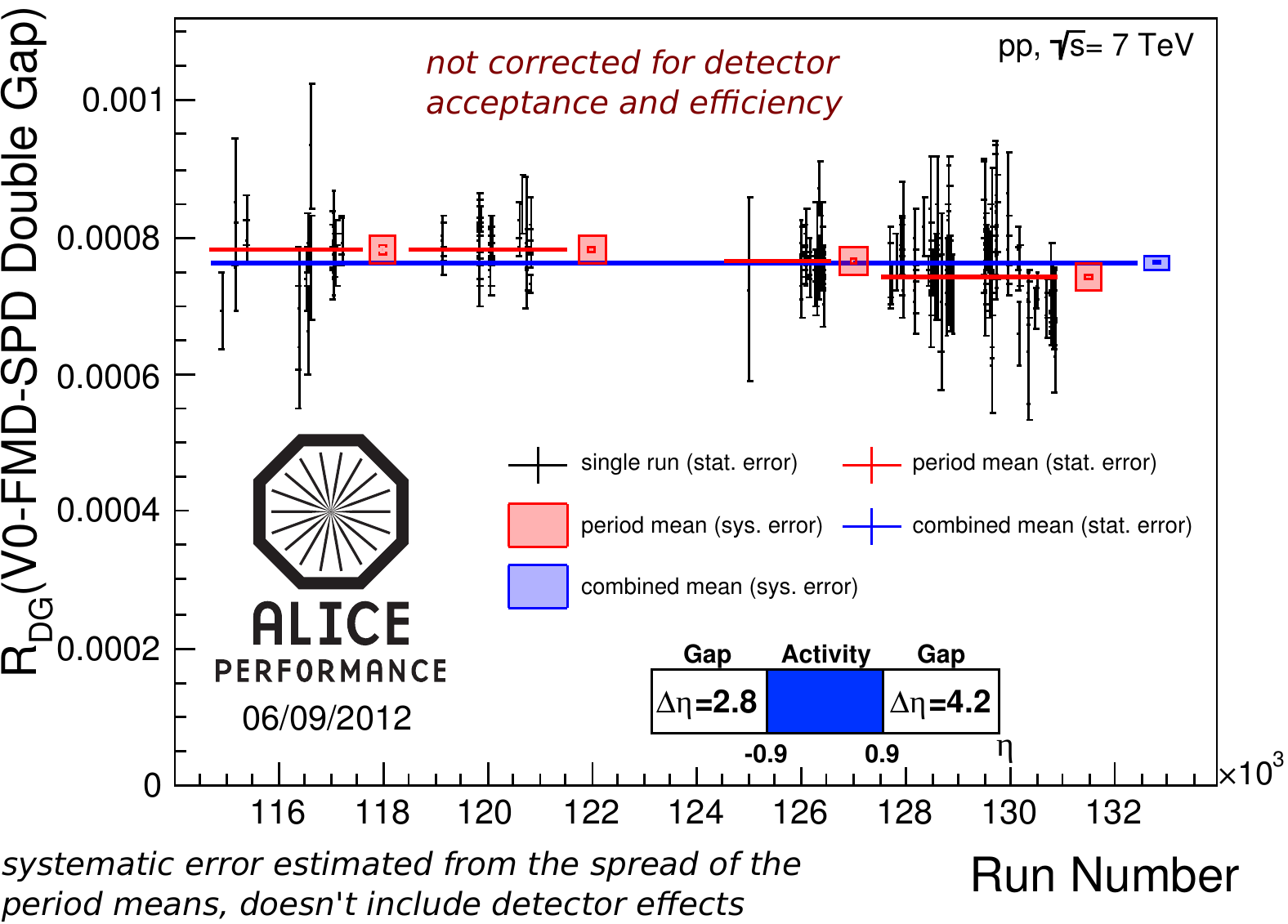}
  \caption{Double-gap fraction in proton-proton collisions at
    $\sqrt{s}=7$\,TeV.}
  \label{fig:dgFraction}
\end{figure}

The double-gap fraction $R_{DG}$ is a measure for the potential amount of
central-diffractive events within the minimum-bias data. In the ALICE
proton-proton data it is found to be
\begin{equation}
  R_{DG}(\underbrace{2.8}_{\Delta\eta_1},\underbrace{\pm0.9}_{\eta_\mathrm{act.}},\underbrace{4.2}_{\Delta\eta_2})=\frac{N_{DG}}{N_\mathrm{MBand}}=(7.63\pm0.02(stat.)\pm0.95(syst.))\cdot 10^{-4}~.
\end{equation}
In Fig.~\ref{fig:dgFraction}, the fraction $R_{DG}$ is shown as a
function of the run number. It shows a very uniform behaviour and no
apparent dependence on the run conditions.

\section{Comparison to Monte~Carlo Generators}
The data simulated using PHOJET and PYTHIA6 were used in order to
estimate the amount of non-diffractive, single and double
diffractive-dissociative interactions having a double-gap
topology. Both PHOJET and PYTHIA6 are used with a tune for single
and double diffractive dissociation but without central
diffraction~\cite{Abelev2012} in order to generate the data set for
comparisons. However, further tuning is needed since generator
multiplicity distributions are too low by up to $30\%$ and $40\%$ in
the central and forward region, respectively. A first estimate of the
bias caused by this deviation assumes the gaps to be uncorrelated and
the multiplicity to be distributed according to a negative binomial
distribution. This estimate leads to a reduction of the double-gap
fraction seen in MC data as shown in
Table~\ref{tab:MCdgf}.
\begin{table}[h]
  \begin{tabular}{lll}
    \hline
    ~                          & $\mathbf{R_{DG}(\mathrm{PYTHIA6})}$  & $\mathbf{R_{DG}(\mathrm{PHOJET})}$   \\ \hline
    not tuned for multiplicity & $3.4\cdot10^{-4}$         & $0.5\cdot10^{-4}$ \\
    first tuning               & $1.0\cdot10^{-4}$         & $0.2\cdot10^{-4}$ \\
    \hline
  \end{tabular}
  \caption{Double-gap fraction in Monte Carlo datasets, before and
    after first tuning}
  \label{tab:MCdgf}
\end{table}
Both generators are unable to describe the double-gap fraction
\mbox{$7.63\cdot 10^{-4}$} observed in ALICE data. These findings can
be interpreted as hints for central diffraction, since the simulated
background amounts only to about $10\%$. However, the ALICE acceptance
and efficiency for central-diffractive events need to be studied using
Monte Carlo data.

\section{Summary and Outlook}
ALICE is well suited to measure central diffraction due to the low
\pt-threshold and the low luminosity data set. The double-gap fraction
derived from data of \mbox{$7.63\pm0.02(stat.)\pm0.87(syst.)\cdot
  10^{-4}$} is in excess by about a factor 10 over the expected
background simulated by PYTHIA6 and PHOJET. The evaluation of
systematic uncertainties will be refined in the future by tuning the
MC simulations of the backgrounds and the gap signal will be
turned into a cross section measurement making the result more
universal. Further studies will determine the properties of this
signal by studying the gap-size dependence of the double-gap fraction
and specific states.


\begin{theacknowledgments}
This work is supported by the German Federal Ministry of Education and
Research under promotional reference 06HD197D and by WP8 of the hadron
physics programme of the 7$^\mathrm{th}$ and 8$^\mathrm{th}$ EU
programme period.
\end{theacknowledgments}



\bibliographystyle{aipproc}   

\bibliography{diffraction2012}

\IfFileExists{\jobname.bbl}{}
 {\typeout{}
  \typeout{******************************************}
  \typeout{** Please run "bibtex \jobname" to optain}
  \typeout{** the bibliography and then re-run LaTeX}
  \typeout{** twice to fix the references!}
  \typeout{******************************************}
  \typeout{}
 }

\end{document}